\documentclass[%
 reprint,
 superscriptaddress,
 amsmath,amssymb,
 aps,
pre,
 floatfix
]{revtex4-2}

\usepackage{lineno}
\usepackage{graphicx}
\usepackage{mathtools}
\usepackage{float}
\usepackage{dcolumn}
\usepackage{bm}
\usepackage{booktabs}
\usepackage{hyperref}
\hypersetup{
    colorlinks=true, 
    linktoc=all,     
    linkcolor=blue,  
    allcolors=blue,
}
\usepackage{array}
\newcolumntype{P}[1]{>{\centering\arraybackslash}p{#1}}

\usepackage{xcolor}
\definecolor{forestgreen}{rgb}{0.33,0.61,0.34}

\begin{document}

\preprint{APS/123-QED}

\title{Metapopulation models imply non-Poissonian statistics of interevent times}

\author{Elohim Fonseca dos Reis}
\affiliation{Department of Mathematics, State University of New York at Buffalo, Buffalo, New York, USA}
\author{Naoki Masuda}
\email{naokimas@buffalo.edu}
\affiliation{Department of Mathematics, State University of New York at Buffalo, Buffalo, New York, USA}
\affiliation{Computational and Data-Enabled Science and Engineering Program, State University of New York at Buffalo, Buffalo, New York, USA}
\affiliation{Faculty of Science and Engineering, Waseda University, Tokyo, Japan}

\date{\today}

\begin{abstract}
Interevent times in temporal contact data from humans and animals typically obey heavy-tailed distributions, which impacts contagion and other dynamical processes on networks.
We theoretically show that distributions of interevent times heavier-tailed than exponential distributions are a consequence of the most basic metapopulation model used in epidemiology and ecology, in which individuals move from one patch to another according to the simple random walk.
Our results hold true irrespective of the network structure and also for more realistic mobility rules such as high-order random walks and the recurrent mobility patterns used for modeling human dynamics.
\end{abstract}

\maketitle


\section{\label{sec:introduction}Introduction}

Contact networks among individuals crucially impact contagion in populations and networks, including spreading of infectious diseases, information, opinions, and rumors.
In addition to the network structure, temporal aspects of contacts and edges in networks, collectively referred to as temporal networks, play a central role in such dynamics \cite{HolmeSaramaki2012PhysRep, Masuda2020Book, Karsai2018book, HolmeSaramaki2019book}.
The interevent time (IET), defined as the time between consecutive events, is one such aspect.
Event-driven stochastic dynamics models on networks most typically assume Poisson processes for generating events, which yield exponential IET distributions.
However, IETs from various human activity data obey non-Poissonian statistics, in particular, heavy-tailed distributions \cite{HolmeSaramaki2012PhysRep, Karsai2018book, Barabasi2005Nature, Vazquez2006PRE}.
In most (but not all) cases, heavy-tailed IET distributions slow down contagion and diffusion in epidemic processes \cite{Min2011PRE, Karsai2011PRE, Rocha2011PlosComputBio, Miritello2011PRE, Masuda2013F1000prime, Jo2014PRX, PastorSatorras2015RMF, Masuda2017book}, opinion dynamics \cite{Wu2010PhysicaA, Takaguchi2011PRE, FernandezGracia2011PRE, Nishi2014EPL}, evolutionary game dynamics \cite{Li2020NatComm}, cascade processes \cite{Karimi2013PhysicaA, Takaguchi2013PLoSOne, Backlund2014PRE, Unicomb2021NatComm}, and random walks \cite{Hoffmann2012PRE, Starnini2012PRE, Speidel2015PRE, Masuda2017PhysRep}.
Several mechanisms can generate heavy-tailed IET distributions, including priority queue models \cite{Barabasi2005Nature, Vazquez2005PRL, Vazquez2006PRE, Grinstein2006PRL, Masuda2009PRE, Oliveira2009PhysicaA, Jo2012PRE}, self-exciting processes \cite{Malmgren2008PNAS, Malmgren2009Science, Masuda2013bookchap}, mixture of exponentials \cite{MasudaHolme2020PRR, Okada2020RSOS, Jiang2016JStatMech}, and dynamics of nodal states, where mutual agreement of two nodes produces contact events at a high rate \cite{Reis2020PRE}.

However, these models do not explain the genesis of heavy-tailed IET distributions in the presence of mobility of individuals.
Human and animal individuals move around to meet others, transmitting information and disease, and then depart from each other.
In fact, most of face-to-face contact or proximity data have been collected from mobile individuals, and such data show heavy-tailed IET distributions \cite{Masuda2011PRX, Panisson2012AdHocNet, Starnini2012PRE, Panisson2013bookchap, Barrat2013bookchap, Gauvin2013SciRep, Fournet2014PlosOne, Vestergaard2014PRE, Genois2018EPJ, Reis2020PRE}.
Previous modeling studies showed that encountering events of random walkers moving on a two-dimensional plane with heterogeneous attractiveness values generate heavy-tailed IET distributions \cite{Starnini2013PRL, Starnini2016SocNet, Zhang2016EuroPhysJB, Starnini2016SciRep, Flores2018PRL}.
These studies crucially assumed that each walker is endowed with a randomly assigned attractiveness value and that walkers tend to slow down when they approach attractive walkers nearby.
In the present study, we are interested in analytically accounting for heavy-tailed IET distributions under mobility in a simpler manner, i.e., without introducing the concept of attractiveness or any heterogeneity among walkers.

A prevalent approach to simultaneous modeling of mobility and agent-to-agent interaction is metapopulation models \cite{Colizza2007NatPhys}, originally introduced in mathematical epidemiology \cite{Hethcote1978TheoPopBiol, May1984MathBiosci, Lloyd1996JTheorBiol, Grenfell1997TrendsEcolEvol, Grenfell1998EcolLett} and ecology \cite{Hanski1998Nature, Hanski1997book, Hanski2004book}.
In a metapopulation network, nodes represent subpopulations, such as, households, cities, urban areas, or ecological habitats, and edges represent migration routes.
Metapopulation models have been successful in describing, e.g., epidemic processes  \cite{Colizza2007PRL, Colizza2008JTheorBiol, Barrat2008book, PastorSatorras2015RMF, Masuda2020Book} including the  COVID-19 pandemic \cite{Chinazzi2020Science}.

In the present study, we provide accounts of heavy-tailed IET distributions when individuals move around, meet, and separate according to a standard metapopulation network model, in which the individuals perform simple random walks, and its more realistic variants.
We analytically show that the IET distribution for a pair of individuals is a mixture of exponential distributions, which robustly holds true for different network structures and mobility rules.
Crucially, mixtures of exponential distributions produce IET distributions that are substantially closer to heavy-tailed distributions observed in empirical data than exponential distributions do.

\section{\label{sec:model}Model} 

We consider a metapopulation network with $N$ subpopulations (i.e., nodes).
The network may be directed and/or weighted.
The individuals that populate the network are assumed to be simple random walkers in continuous time.
Different walkers move independently of each other and may produce contact events (events for short) between them only when they are copresent in the same subpopulation.

\begin{figure*}[ht]
\includegraphics{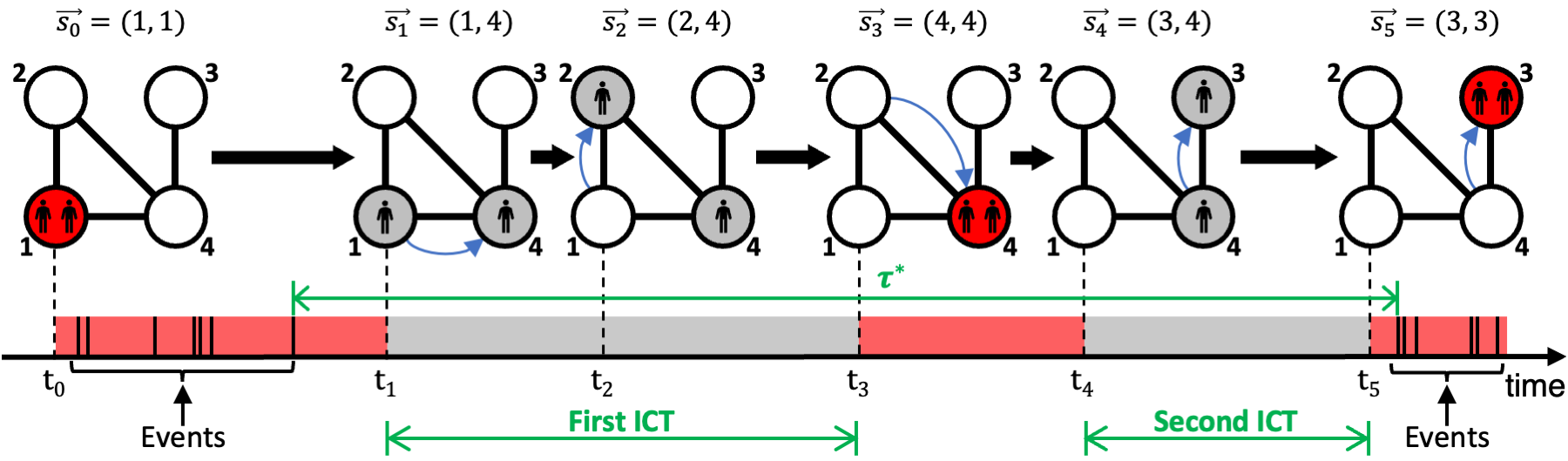}
\caption{\label{fig:dynamics_schematic}
	Schematic illustration of the dynamics of two walkers in a network with four subpopulations.
	Initially, at $t=t_0$, the walkers are copresent in subpopulation $1$.
	Then, events between them occur before $t_1$.
	At $t=t_1$, one walker moves from subpopulation 1 to subpopulation 4.
	At $t=t_2$, the other walker moves from subpopulation 1 to subpopulation 2.
	At $t=t_3$, the same walker moves to subpopulation 4 to be copresent with the other walker.
	Therefore, the first ICT is given by $t_3 - t_1$.
	Although the walkers are copresent for $t \in (t_3, t_4)$, no event occurs.
	At $t=t_4$, one walker moves to subpopulation 3.
	At $t=t_5$, the other walker also moves to subpopulation 3, producing a second ICT, which is equal to $t_5-t_4$.
	Finally, an event occurs during their copresence in subpopulation 3, yielding an IET, denoted by $\tau^*$.
	In this example, this IET is followed by several much shorter IETs.
	}
\end{figure*}

Specifically, the time $t$ for which each walker waits until it leaves the current subpopulation is independently distributed according to an exponential distribution with rate $\mathcal{D}$.
The mobility and contact dynamics are illustrated in Fig.~\ref{fig:dynamics_schematic}.
When the walker leaves the $i$th subpopulation, it moves to a neighboring $j$th subpopulation with probability $A_{ij}/\sum_{\ell=1}^N A_{i\ell}$, where $A_{ij}$ is the weight of edge $(i, j)$.
Events between a pair of walkers in the same subpopulation occur according to a Poisson process with rate $\lambda$, which epidemic process models on metapopulation networks usually assume.
In other words, the distribution of time $t$ to the next event between a pair of walkers in the same subpopulation is given by $\phi_{\rm e}(t) = \lambda e^{-\lambda t}$.
When the walkers are in different subpopulations, no event occurs between them.

\section{Results} \label{sec:results}

\subsection{General solution} \label{sec:general_solution}

Consider two walkers visiting the same subpopulation.
We derive the probability density of an IET, $\tau$, between an arbitrary pair of individuals, which are random walkers.
When the two walkers are copresent in a subpopulation, events (including the case of 0 event) occur between them before either walker leaves the subpopulation.
After that, a series of $n$ copresences occurs before the next event occurs (see Fig.\ref{fig:dynamics_schematic}).

We let $p(\tau, n)$ be the probability of generating an IET of length $\tau$ during the $n$th copresence after either walker leaves the current subpopulation.
The probability density of the time at which either walker leaves the subpopulation where the two walkers are copresent is given by $\phi_{\rm d}(t) = 2\mathcal{D} e^{-2\mathcal{D} t}$.
We obtain
\begin{eqnarray}
p(\tau, 0) = \int_{\tau}^\infty dt^\prime \phi_{\rm d}(t^\prime) \int_{\tau}^{\infty} dt^{\prime\prime} \phi_{\rm e}(t^{\prime\prime}) \lambda = \lambda e^{-(\lambda+2\mathcal{D})\tau}.\quad
\end{eqnarray}
By Laplace transforming $p(\tau, 0)$, we obtain 
\begin{eqnarray}
\hat{p}(s,0) \equiv \int_0^\infty d \tau e^{-s\tau} p(\tau, 0) = \frac{\lambda}{(s+\lambda+2\mathcal{D})}.
\end{eqnarray}

We denote the distribution of time $t$ between consecutive copresences of the two walkers, which we refer to as the inter-copresence time (ICT), by $\phi_{\rm c}(t)$.
Then, the joint probability density of $\tau$ and $n=1$ is given by
\begin{align}
p(\tau, 1) &= \int_0^\tau d t_1 S_{\rm e}(t_1) S_{\rm d}(t_1) 2\mathcal{D} \int_{t_1}^\tau d t_2 S_{\rm c}(t_2-t_1) \nonumber \\
	& \times h_{\rm c}(t_2-t_1) S_{\rm d}(\tau-t_2)S_{\rm e}(\tau-t_2) \lambda \nonumber \\
	&= 2\mathcal{D}\lambda \int_0^\tau d t_1 \int_{t_1}^\tau d t_2 S_{\rm d}(t_1)S_{\rm e}(t_1) \phi_{\rm c}(t_2-t_1) \nonumber \\
	& \times S_{\rm d}(\tau-t_2)S_{\rm e}(\tau-t_2)\,.
\end{align}
where $S_{\rm e}(t)=e^{-\lambda t}$ and $S_{\rm d}(t)=e^{-2\mathcal{D}t}$ are the survival functions of $\phi_{\rm e}(t)$ and $\phi_{\rm d}(t)$, respectively, and $S_{\rm c}(t)=\int_{t}^\infty dt^\prime \phi_{\rm c}(t^\prime)$ and $h_{\rm c}(t)=\phi_{\rm c}(t)/S_{\rm c}(t)$ are the survival and hazard functions of $\phi_{\rm c}(t)$, respectively.
By Laplace transforming $p(\tau, 1)$, we obtain
\begin{eqnarray}
\hat{p}(s, 1) = \frac{2\lambda \mathcal{D} \hat{\phi}_{\rm c}(s)}{(s+\lambda+2\mathcal{D})^2}.
\end{eqnarray}
Similarly, the Laplace transform of the joint probability density of $\tau$ and $n$ is given by
\begin{eqnarray} \label{eq:p(s,n)}
\hat{p}(s, n) = \frac{\lambda}{s+\lambda+2\mathcal{D}} \left(\frac{2\mathcal{D} \hat{\phi}_{\rm c}(s)}{s+\lambda+2\mathcal{D}}\right)^n.
\end{eqnarray}
Therefore, the IET distribution in the frequency domain is given by
\begin{eqnarray} \label{eq:general_IET_dist}
\hat{p}(s) = \sum_{n=0}^\infty \hat{p}(s, n) = \frac{\lambda}{s + \lambda + 2\mathcal{D}[1-\hat{\phi}_{\rm c}(s)]}\,.
\end{eqnarray}

Using Eq.~\eqref{eq:general_IET_dist}, we obtain the following expression of the coefficient of variation (CV), defined as the standard deviation divided by the mean, of $\tau$ (see Appendix \ref{appendix:CV} for the derivation):
\begin{eqnarray}\label{eq:CV_general}
\mathrm{CV} = \sqrt{1 + \frac{2\lambda \mathcal{D} \hat{\phi}^{\prime\prime}_{\rm c}(0) }{[1-2\mathcal{D}\hat{\phi}^{\prime}_{\rm c}(0)]^2}}\,,
\end{eqnarray}
where $\hat{\phi}^{\prime}_{\rm c}(0)$ and $\hat{\phi}^{\prime\prime}_{\rm c}(0)$ are the first and second derivatives of $\hat{\phi}_{\rm c}(s)$ evaluated at $s=0$, respectively.

\subsection{Exponential ansatz} \label{sec:ansatz}

The asymptotic form of the distribution of first-passage time of random walks in networks with short relaxation time has an exponential tail \cite{Lau2010EPL}.
With $\phi_{\rm c}(t) = \alpha e^{-\alpha t}$, which we refer to as the exponential ansatz, Eq.~\eqref{eq:general_IET_dist} is reduced to
\begin{eqnarray} \label{eq:ansatz_IET_dist_s}
\hat{p}(s) = \frac{\lambda(s+\alpha)}{(s+\alpha)(s+\lambda+2\mathcal{D})-2\mathcal{D}\alpha}\,.
\end{eqnarray}
The inverse Laplace transform of Eq.~\eqref{eq:ansatz_IET_dist_s} is given by
\begin{align} \label{eq:ansatz_IET_dist}
p(\tau) =& \frac{\lambda}{2 \Delta} \left[ (\Delta-\lambda-2\mathcal{D}+\alpha) e^{-\frac{(\lambda+2\mathcal{D}+\alpha-\Delta)\tau}{2}} \right. \nonumber \\
	&\left.+ (\Delta+\lambda+2\mathcal{D}-\alpha) e^{-\frac{(\lambda+2\mathcal{D}+\alpha+\Delta)\tau}{2}} \right],
\end{align}
where $\Delta \equiv \sqrt{(\lambda+2\mathcal{D}+\alpha)^2-4\alpha\lambda} > 0$.
Therefore, the distribution of IET is a mixture of two exponential distributions (see Appendix \ref{appendix:ansatz_dist} for a proof).
By substituting $\phi_{\rm c}(t) = \alpha e^{-\alpha t}$ in Eq.~\eqref{eq:CV_general}, we obtain
\begin{eqnarray}\label{eq:CV_ansatz}
{\rm CV_{ansatz}} = \sqrt{1+\frac{4\lambda\mathcal{D}}{(2\mathcal{D}+\alpha)^2}}\,.
\end{eqnarray}

\subsection{\label{sec:mixture_solution}Intercopresence time distribution of the walkers}

To mathematically derive the distribution of ICTs, $\phi_{\rm c}(t)$, we denote by $\vec{s}=(m,n)$ the state of the system in which one  walker is in the $m$th subpopulation and the other walker is in the $n$th subpopulation.
Because the walkers are indistinguishable, the system has $\overline{N}=N(N+1)/2$ states, i.e., $(m,n)$, where $1 \leq m \leq n \leq N$.

We consider the first-passage time of a continuous-time Markov chain from a state $(i, i)$, i.e., a copresence, to a state $(j,j)$, i.e., the next copresence, where $i, j \in \{1, 2, \ldots, N\}$.
A similar problem of first-passage time to the next copresence of two random walkers appears in the theory of cooperation in evolutionary games on networks \cite{Allen2017Nature, Fotouhi2019JRSocInterface}.
We refer to the states in which the two walkers are in the same subpopulation as absorbing states and those in which the walkers are in different subpopulations as transient states.
Let $\mathcal{A} = \{(m,n);m=n\}$  and $\mathcal{B}=\{(m,n);m<n\}$ be the set of absorbing and transient states, respectively.
We denote the probability that the dynamics starting in state $(i,i)$ at time $0$ is in state $\vec{s}=(m,n)$, where $m \leq n$, at time $t$ by $p_{i,\vec{s}}(t)$, or equivalently, $p_{i,(m,n)}(t)$.
We set 
\begin{eqnarray} \label{eq:pTA}
{\bm p}_i(t)=\begin{bmatrix} {\bm p}_i^T(t) & {\bm p}_i^A(t) \end{bmatrix},
\end{eqnarray}
where ${\bm p}_i^T(t)= [ p_{i,(1,2)}(t), p_{i,(1,3)}(t), \ldots, p_{i,(N-1, N)}(t)]$ is the $N_T \equiv N(N-1)/2$-dimensional vector of the transient states, and ${\bm p}_i^A(t)=[p_{i,(1,1)}(t), p_{i,(2,2)}(t), \ldots, p_{i,(N, N)}(t)]$ is the $N$-dimensional vector of absorbing states.

At time 0, one of the walkers leaves subpopulation $i$ such that the system leaves the absorbing state $(i, i)$ and enters a transient state.
Therefore, the distribution of the initial state of the Markov process is given by 
\begin{eqnarray} \label{eq:initial_state}
p_{i,(m,n)}(0)=(1-\delta_{mn})\frac{A_{in} \delta_{mi} + A_{im} \delta_{ni}}{k^{\rm out}_i},
\end{eqnarray}
where $k^{\rm out}_i = \sum_{l=1}^N A_{il}$ is the out-degree of subpopulation $i$, and $\delta_{ij}$ is the Kronecker delta.

The master equation for a transient state $\vec{s}=(m,n)$ is given by
\begin{eqnarray} \label{eq:master_pTi}
\frac{d \bm p_i^T(t)}{dt} = - \bm p_i^T(t) L\,,
\end{eqnarray}
where $L \equiv 2\mathcal{D}I-Q$, matrix $Q$ is the matrix of transitions rates between transient states, and $I$ is the $N_T \times N_T$ identity matrix (see Appendix \ref{appendix:transition_rate_matrix} for details).
Equation \eqref{eq:master_pTi} leads to
\begin{eqnarray} \label{eq:pT}
\bm p_i^T(t) = \bm p_i^T(0) e^{-Lt}\,.
\end{eqnarray}

The master equation for the absorbing states is given by
\begin{eqnarray} \label{eq:master_pA}
\frac{d \bm p_i^A(t)}{d t} =  \bm p_i^T(t) R\,,
\end{eqnarray}
where $R$ is the matrix of transition rates from a transient state to an absorbing state (see Appendix \ref{appendix:transition_rate_matrix}).
By substituting Eq.~\eqref{eq:pT} into Eq.~\eqref{eq:master_pA}, we obtain
\begin{eqnarray}
\bm p_i^A(t) = \bm p_i^T(0) L^{-1} (I-e^{-Lt}) R\,.
\end{eqnarray}

The probability density function of the first-passage time from $(i,i)$ to a state $\vec{s}$, denoted by $f_{i,\vec{s}}(t)$, satisfies
\begin{eqnarray}
    p_{i, \vec{s}}(t) = \delta(t) \delta_{i,\vec{s}} + \int_0^t f_{i, \vec{s}}(t^{\prime}) p_{\vec{s},\vec{s}}(t-t^{\prime}) dt^{\prime},
\end{eqnarray}
where $\delta_{i, \vec{s}}=1$ if $\vec{s}=(i,i)$; otherwise $\delta_{i, \vec{s}}=0$.
For $\vec{s} \in \mathcal{A}$, we obtain
\begin{eqnarray}
    p_{i, \vec{s}}^A(t) = \int_0^t f_{i, \vec{s}}^A(t^{\prime}) dt^{\prime},
\end{eqnarray}
where $f_{i, \vec{s}}^A(t)$, or equivalently $f_{ij}^A(t)$, is the probability density of the first-passage time from $(i, i)$ to $(j, j)$, because $p_{\vec{s}, \vec{s}}(t-t^\prime) = 1$ for any $t \geq t^\prime$.
Therefore, the probability density of the first-passage time to the absorbing states is given by
\begin{eqnarray} \label{eq:f_i^A(t)}
\bm f_i^A(t) = \frac{d \bm p_i^A(t)}{d t} = \bm p_i^T(0) e^{-Lt} R\,,
\end{eqnarray}
where $\bm f_i^A(t) = [f_{i1}^A(t), \ldots, f_{iN}^A(t)]$, and $f_{ij}^A(t)$ is the probability density of the first-passage time from $(i,i)$ to $(j,j)$.

We define the $N \times N_T$ matrix $P^T(t)$ such that its $i$th row is given by $\bm p^T_i(t)$, and the $N \times N$ matrix $F^A(t)$ by $[F^A(t)]_{ij} = f_{ij}^A(t)$.
Then, Eq.~\eqref{eq:f_i^A(t)} is equivalent to 
\begin{eqnarray} \label{eq:matrix_fpt_abs}
F^A(t) = P^T_0 e^{-Lt} R\,,
\end{eqnarray}
where $P^T_0 = P^T(0)$.

We are interested in the IET distribution in the equilibrium.
In the equilibrium, the initial copresence state $(i,i)$ should be the stationary probability of the discrete-time random walk on $\mathcal{A}$, where the transition probability $T_{ij}$ from $(i,i)$ to $(j,j)$ is given by
\begin{eqnarray} \label{eq:T_coeff}
T_{ij} = \int_0^\infty dt f_{ij}^A(t)\,.
\end{eqnarray}
Using Eqs.~\eqref{eq:matrix_fpt_abs} and \eqref{eq:T_coeff}, we obtain the $N \times N$ matrix $[T]_{ij}=T_{ij}$ as
\begin{eqnarray} \label{eq:T_matrix}
T = P^T_0 L^{-1} R\,.
\end{eqnarray}
We obtain the stationary distribution $\bm q^*$ for the absorbing states as
\begin{eqnarray} \label{eq:stationary_abs}
\bm q^* T = \bm q^*\,,
\end{eqnarray}
which we can numerically solve for arbitrary networks of subpopulations.

If $Q$ is diagonalizable, we can write
\begin{eqnarray} \label{eq:diagonal_eQt}
e^{Qt} =  \sum_{j=1}^{N_T} e^{\gamma_j t} \bm v_j \bm w_j\,,
\end{eqnarray}
where $\bm v_j$ and $\bm w_j$ are the right and left eigenvectors, respectively, associated with eigenvalue $\gamma_j$ of $Q$.

By combining Eqs.~\eqref{eq:matrix_fpt_abs}, \eqref{eq:stationary_abs}, and \eqref{eq:diagonal_eQt}, we obtain the ICT distribution weighted by the stationary probability of the initial location of the two copresent walkers as
\begin{eqnarray} \label{eq:ICT_mix_exp_dist}
\phi_{\rm c}(t) = \bm q^* F^A(t) \bm 1_N = \sum_{j=1}^{N_T} c_j \alpha_j e^{-\alpha_j t},
\end{eqnarray}
where $c_j = \bm q^* P^T_0 \bm v_j \bm w_j \bm 1_{N_T}$, $\alpha_j = 2\mathcal{D}-\gamma_j > 0$, and ${\bm1}_{N}$ and ${\bm1}_{N_T}$ are the column vectors of size $N$ and $N_T$, respectively, in which all the elements are 1 [see Appendix \ref{appendix:dist_ICT} for a proof of Eq.~\eqref{eq:ICT_mix_exp_dist} and that $\alpha_j>0$].
Because $\sum_{j=1}^{N_T} \bm v_j \bm w_j = I$, we have $\sum_{j=1}^{N_T} c_j = 1$.
Therefore, $\phi_{\rm c}(t)$ is a mixture of exponential distributions, including the case in which $c_j < 0$.

By substituting the Laplace transform of Eq.~\eqref{eq:ICT_mix_exp_dist} into Eq.~\eqref{eq:general_IET_dist} and Eq.~\eqref{eq:CV_general}, we obtain the IET distribution in the frequency domain and the CV as
\begin{eqnarray} \label{eq:mixture_IET_dist_s}
\hat{p}(s) = \frac{\lambda}{s+\lambda+2\mathcal{D}[1 - \bm q^* P_0^T (sI+L)^{-1}R \bm 1_N]}
\end{eqnarray}
and
\begin{eqnarray}\label{eq:CV_mixture}
{\rm CV_{mixture}} = \sqrt{1 + \frac{4 \lambda \mathcal{D} \bm q^* P_0^T L^{-1} L^{-1} \bm 1_{N_T}}{(1+ 2 \mathcal{D} \bm q ^* P_0^T L^{-1} \bm 1_{N_T})^2}}\,,
\end{eqnarray}
respectively.
This solution depends on the network structure through $\phi_{\rm c}(t)$.

To compare ${\rm CV_{ansatz}}$ with ${\rm CV_{mixture}}$ we impose that the mean ICT is equal between the two.
This condition combined with $\phi_c(t) = \alpha e^{-\alpha t}$ and Eq.~\eqref{eq:ICT_mix_exp_dist} yields 
\begin{eqnarray}
1/\alpha = \sum_{j=1}^{N_T} (c_j/\alpha_j).
\end{eqnarray}
Then, ${\rm CV_{mixture}}$ is larger than ${\rm CV_{ansatz}}$ because $\sum_m (c_m/\alpha_m^2) > \left[ \sum_m (c_m/\alpha_m)\right]^2$, which is satisfied by the Sedrakyan's inequality (also known as Titu's lemma) \cite{Sedrakyan2018book}.
Therefore, the exponential ansatz, which has led to the mixture of two exponential distributions for IETs, gives a lower bound in terms of the dispersion of IETs.
Additionally, in the case of the network of two subpopulations connected to each other, Eq.~\eqref{eq:ICT_mix_exp_dist} is reduced to the exponential ansatz (see Appendix \ref{appendix:dist_ICT}).
Therefore, although the time-domain solution of the IET distribution is not available in general, the IETs are guaranteed to be distributed according to a distribution with a heavier tail than a mixture of two exponential distributions.

In fact, if the system starts in state $(i,i)$ and reaches $(j,j)$ to produce an ICT, it restarts from $(j, j)$ to produce the next ICT.
Therefore, Eq.~\eqref{eq:mixture_IET_dist_s}, which assumes the independence of different ICTs, is only approximate.
Nevertheless, the following numerical simulations support that the difference between the approximate solution [i.e., Eqs.~\eqref{eq:mixture_IET_dist_s} and \eqref{eq:CV_mixture}] and the exact solution (see Appendix \ref{appendix:full_solution}) is negligible.

\subsection{\label{sec:numerical_simulaitons}Numerical results}

\begin{figure}[t]
\includegraphics{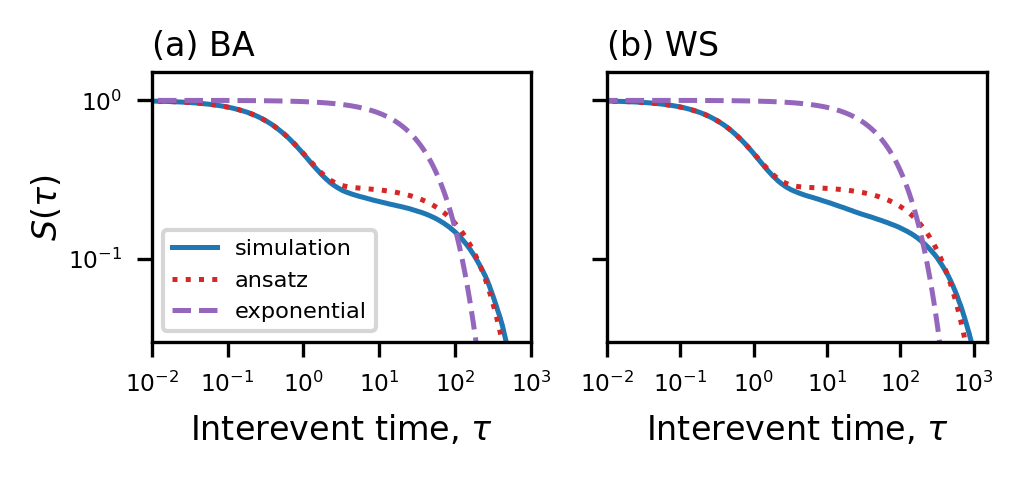}
\caption{\label{fig:IET_distribution}
	Survival function of IETs, $S(\tau)$.
	We show the results for the direct numerical simulation of the model (simulation), the exponential ansatz [ansatz; Eq.~\eqref{eq:ansatz_IET_dist}], and the single exponential distribution (exponential) whose mean is equal to that for the direct simulations.
	In Eq.~\eqref{eq:CV_ansatz}, we set $\alpha = 1/\sum_{j=1}^{N_T} (c_j/\alpha_j) = 1/\bm q^* P_0^T L^{-1} \bm 1_{N_T}$.
	(a) BA network with $m=2$.
	(b) WS network with $p=0.1$.
	We set $N=100$, $\lambda = 1$, and $\mathcal{D} = 0.2$.
	We stop the simulation when $10^5$ events are generated.
	}
\end{figure}

We simulated the model to validate our theory.
We show the IET distribution produced by two walkers on the Barab{\'a}si-Albert (BA) and Watt-Strogatz (WS) networks in Figs.~\ref{fig:IET_distribution}(a) and \ref{fig:IET_distribution}(b), respectively.
The figure shows that the IET distributions obtained from numerical simulations and ansatz have heavier tails than the exponential distribution whose mean IET is equal to that for the numerical simulations.
The distribution obtained from the simulation decays more smoothly than the exponential ansatz.

To be more quantitative and general in terms of the parameter values and variety of networks, we compared the numerically and theoretically obtained CV of IET on six networks.
Simultaneously changing $\lambda$ and $\mathcal{D}$ to $c\lambda$ and $c\mathcal{D}$, where $c>0$, is equivalent to changing the time from $t$ to $ct$ and using the original $\lambda$ and $\mathcal{D}$ values.
This observation is consistent with our theoretical results for the CV [i.e., Eqs.~\eqref{eq:CV_ansatz} and \eqref{eq:CV_mixture}], which only depends on $\mathcal{D}/\lambda$. 
Therefore, we set $\lambda=1$ without loss of generality and varied $\mathcal{D}$.

\begin{figure}[t]
\includegraphics{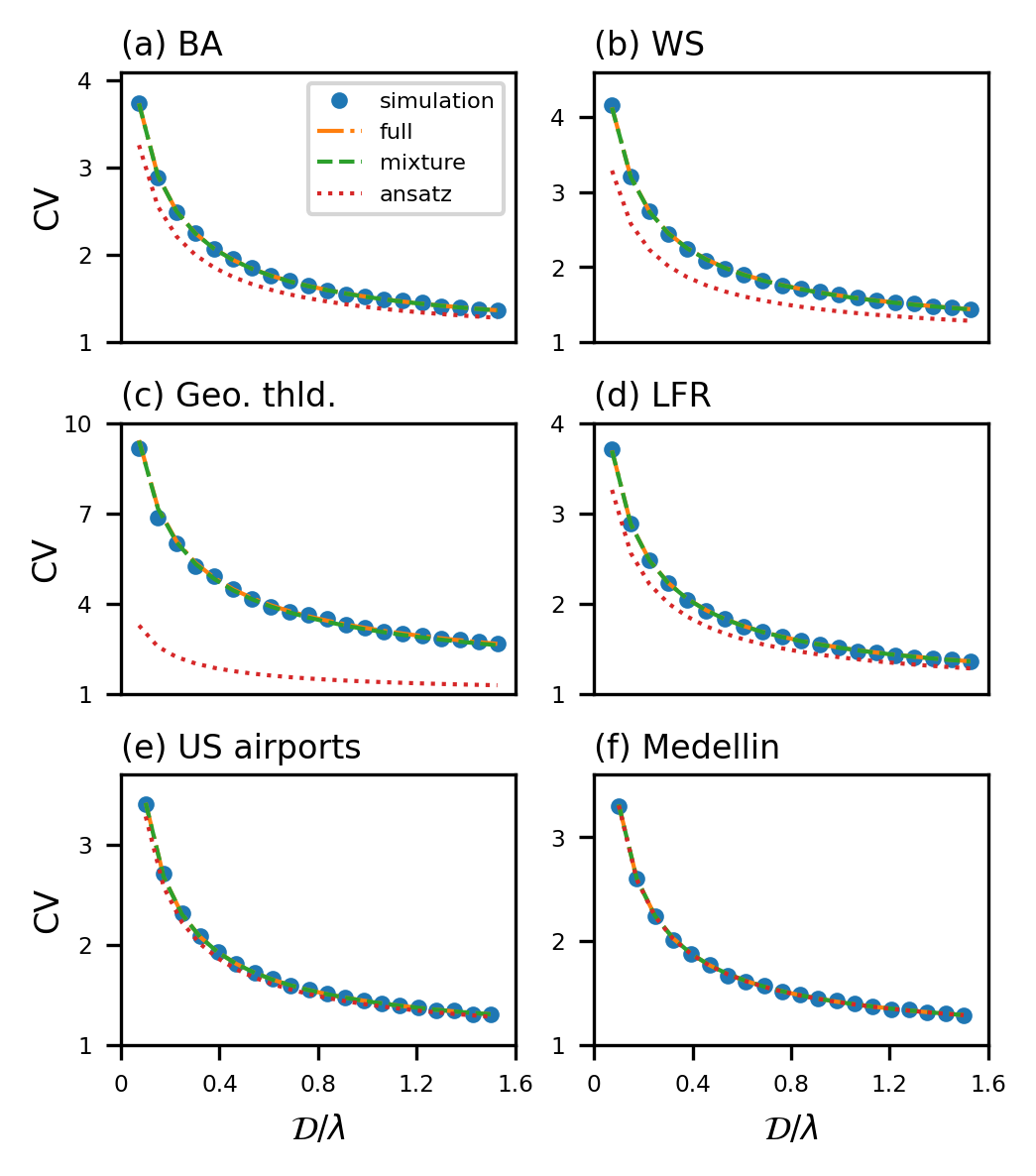}
\caption{\label{fig:CV_curves}
	CV of IET for various metapopulation networks.
	We compare among the simulation of the model (simulation), the exact theoretical solution (full; see Appendix \ref{appendix:full_solution}), our main theory (mixture; Eq.~\eqref{eq:CV_mixture}), and the exponential ansatz (ansatz; Eq.~\eqref{eq:CV_ansatz}).
	(a) Barab{\'a}si-Albert (BA) network with $m=2$ \cite{Barabasi1999Science}.
	(b) Watt-Strogatz (WS) network with $p=0.1$ \cite{Watts1998Nature}.
	(c) Geographical threshold (Geo. thld.) graph with the threshold value $\theta=95$, dimension equal to 2, Euclidean distance metric, and $h(r) = r^{-2}$ \cite{Masuda2005PRE}.
	(d) Lancichinetti–Fortunato–Radicchi (LFR) model with $\gamma=3$, $\beta=1.5$, $\mu=0.2$, average degree equal to 5, maximum degree equal to 50, and minimum community size equal to 10 \cite{Lancichinetti2008PRE}.
	(e) US airport network \cite{USAirport1997}.
	(f) Medellin intercity zone network \cite{Lotero2016RSOS}.
	In panels (a)--(d), the networks have $N=100$ nodes.
	We set $\lambda=1$.	
	}
\end{figure}

The results shown in Fig.~\ref{fig:CV_curves} indicate that the CV is large when $\mathcal{D}/\lambda$ is small for all four model networks and two empirical networks.
Moreover, the CV obtained from our theory [Eq.~\eqref{eq:CV_mixture} and its fuller version derived in Appendix \ref{appendix:full_solution}] is in an excellent agreement with the numerical results.
The exponential ansatz [Eq.~\eqref{eq:CV_ansatz}] is accurate for the two empirical networks [Figs.~\ref{fig:CV_curves}(e) and \ref{fig:CV_curves}(f)), reasonably good for two model networks [Figs.~\ref{fig:CV_curves}(a) and \ref{fig:CV_curves}(d)], and not accurate for the WS and the geographical threshold networks [see Figs.~\ref{fig:CV_curves}(b) and \ref{fig:CV_curves}(c)].
Therefore, spatiality, or the large average path length between nodes, which is present in the last two networks but not in the other networks, may negatively affect the accuracy of the exponential ansatz.

Next, we investigated the effects of the network size on the IET.
We set $\lambda=1$ and $\mathcal{D}=0.25$, and computed the CV value for each of the six networks shown in Fig.~\ref{fig:CV_curves} for different numbers of nodes, i.e., $N=50, 100,200, 500, 1000,$ and $2000$.
We used the same parameter values as those used in Fig.~\ref{fig:CV_curves} for each network model.
\begin{figure}[t]
\includegraphics{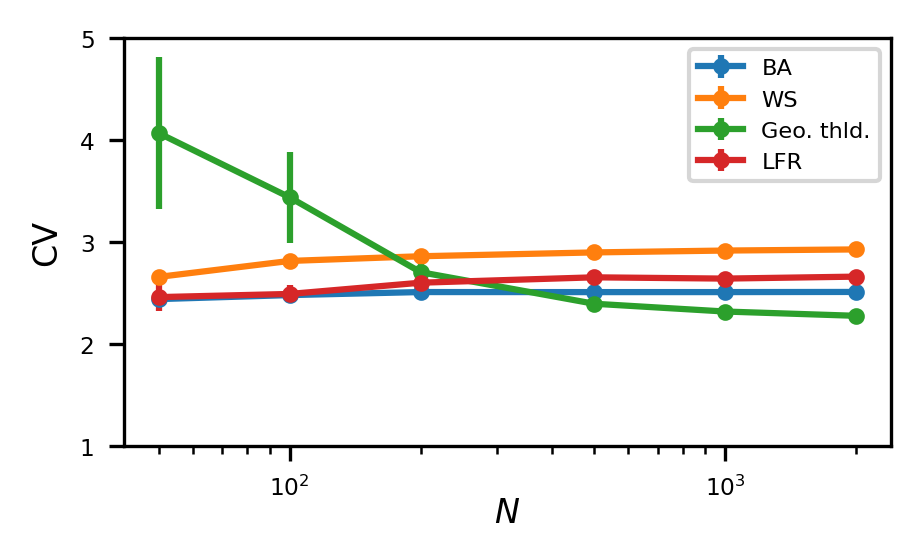}
\caption{\label{fig:size_analysis}
	CV of IET for different networks with different numbers of nodes, i.e., $N=50, 100, 200, 500, 1000, \text{ and } 2000$.
	See the caption of Fig.~\ref{fig:CV_curves} for the names of network models and their parameter values.
	We set  $\lambda=1$ and $\mathcal{D} = 0.25$.
	The error bars represent the standard deviation.
	We computed the average and standard deviation of the CV over ten simulations.
	}
\end{figure}

The CV values for the different network models and different $N$ values are shown in Fig.~\ref{fig:size_analysis}.
We find that the results are roughly independent of $N$, except for the geographical threshold graph.
For the geographical threshold graph, the CV is large at small $N$.
Nevertheless, the CV value for the geographical threshold graph depends little on $N$ for larger networks.

We have assumed simple random walks.
Second-order random walks better approximate human and animal mobility  \cite{Scholtes2014NatComm, Rosvall2014NatComm}.
Our theory and the result that IETs obey a heavier-tailed distribution than the exponential distribution holds true for second- and higher-order random walks (see Appendix \ref{appendix:2nd_order_rw}).

\subsection{\label{sec:work-home_model}Work-home model}

More realistic metapopulation models incorporate the recurrent nature of human mobility where individuals iterate between a home subpopulation and a work subpopulation, which are different for different individuals \cite{Balcan2011NatPhys, Belik2011PRX, Poletto2013JTheorBiol, Gomez2018NatPhys, Granell2018PRE, Soriano2020JStatMech}.
In this section, we analyze the IET distribution for such a mobility rule.

We emulate this situation by assigning to each walker a home subpopulation and allowing it to visit subpopulations adjacent to the home subpopulation.
Then, each walker is confined to a star graph, of which the central subpopulation is the home subpopulation and the leaves are other locations such as work.
Without loss of generality, we consider two random walkers on the respective star graphs to generate sequences of IETs.

\begin{figure*}[ht]
\includegraphics{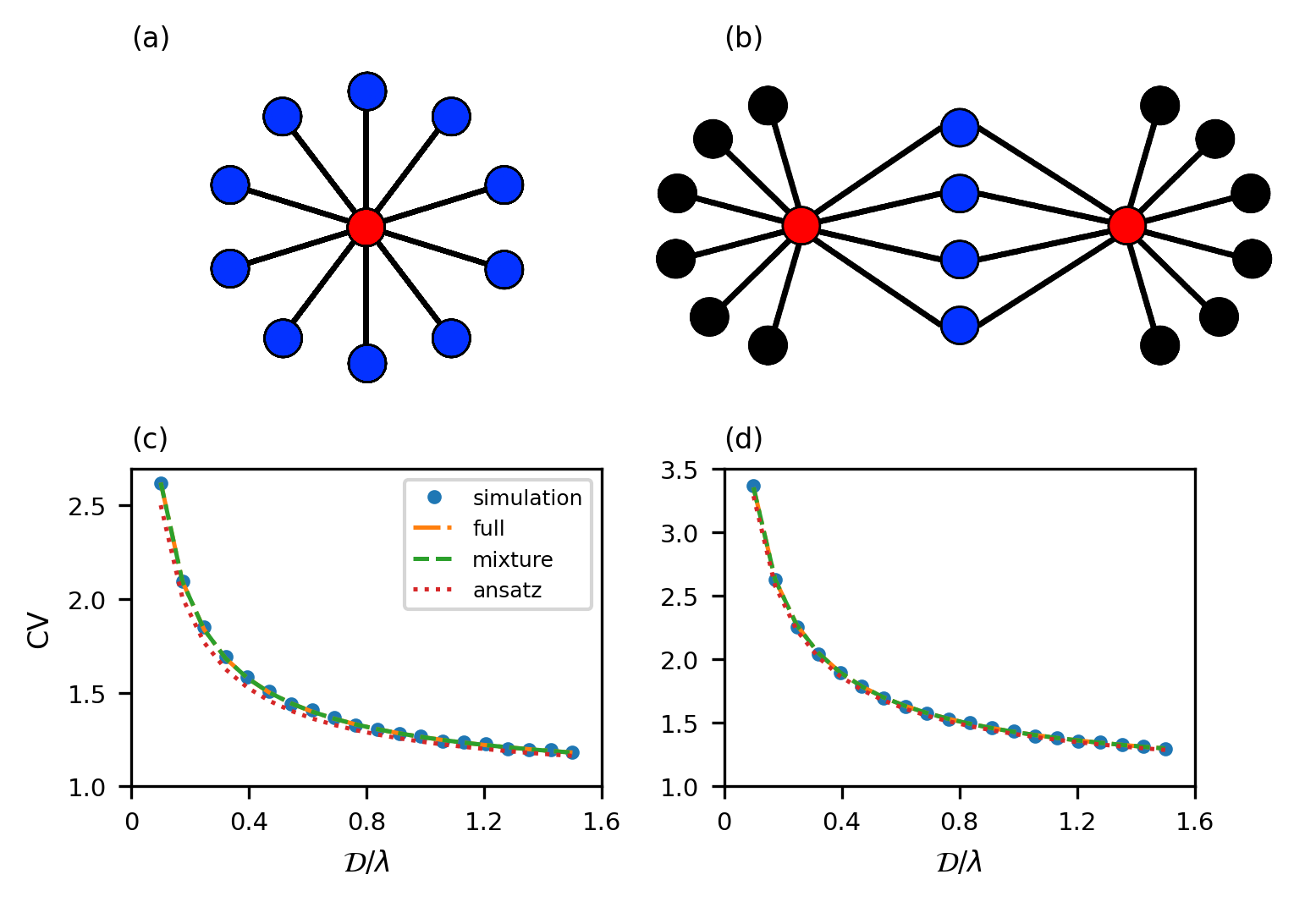}
\caption{\label{fig:work-home_model}
	CV of IET under recurrent mobility patterns.
	Two situations that mimic the recurrent human mobility pattern of commuting between home and other places.
	In the star graph shown in panel (a), the two walkers share the home subpopulation, shown in red, and all the leaf subpopulations, shown in blue.
	In the situation shown in panel (b), the walkers have different home subpopulations and share some leaf subpopulations (shown in blue) but not others (shown in black). 
	We compare the CV of IET among the simulation of the model (simulation), the exact solution (full) from Eq.~\eqref{eq:CV_full}, the mixture of exponential distribution solution (mixture) from Eq.~(\ref{eq:CV_mixture}), and the exponential ansatz solution (ansatz) from Eq.~(\ref{eq:CV_ansatz}) in (c) and (d) for the metapopulation network and the mobility rule shown in (a) and (b), respectively.
	We set $\lambda=1$.
	}
\end{figure*}

We examine two cases.
In the first case, the two walkers share the home subpopulation and all leaf subpopulations, as shown in Fig.~\ref{fig:work-home_model}(a).
In other words, the two walkers are confined to the same star graph, and events between them may occur in any subpopulation.
Therefore, our theory directly applies.

In the second case, we assume that the two walkers have different home subpopulations and share some but not necessarily all the leaf subpopulations, as shown in Fig.~\ref{fig:work-home_model}(b).
This case mimics, for example, the situation in which the two individuals are coworkers living in different cities.
Events may occur between the walkers only when they are copresent in any of the shared leaf subpopulations.
To derive the states and the transition rates between the states, we distinguish the two walkers and the subpopulations.
There are three types of subpopulations: the home subpopulations, which are unshared, the shared leaf subpopulations, which are work locations, for example, and the unshared leaf subpopulations, which represent other locations that one but not both walkers visits.
In a network with $N$ subpopulations, two subpopulations are home subpopulations, $N_s$ subpopulations are the shared leaf subpopulations, and walkers 1 and 2 have $N_1$ and $N_2$ unshared leaf subpopulations, respectively, such that $N=2+N_s+N_1+N_2$.

We define the home subpopulation nodes by $h_1$ and $h_2$, and the set of shared leaf subpopulations by $\mathcal{W}=\{w_1, w_2, \ldots, w_{N_s}\}$.
There are four types of transient states.
In the first type of transient states, both walkers are in different shared leaf subpopulations.
There are $N_s(N_s-1)$ such states.
In the second type, walker 1 is in a shared leaf subpopulation and walker 2 is not
There are $N_s(N_1+1)$ such states.
In the third type, walker 2 is in a shared leaf subpopulation and walker 1 is not
There are $N_s(N_2+1)$ such states.
In the fourth type, both walkers are not in any of the shared leaf subpopulations.
There are $(N_1+1)(N_2+1)$ such states.
Therefore, there are 
\begin{align}
N_T =& N_s(N_s-1) + N_s(N_1+N_2+2) \nonumber \\
     &+ (N_1+1)(N_2+1)
\end{align}
transient states in total.
Each element of the $N_T \times N_T$ matrix of transition rates between transient states, $Q_{\vec{s}\,\vec{s}^{\,\prime}}$, where $\vec{s} = (m_1, m_2) \in \mathcal{B}$ and $\vec{s}^{\,\prime} = (m_1^\prime, m_2^\prime) \in \mathcal{B}$, is given by

\begin{widetext}
\begin{align} \label{eq:Q_diff_home}
Q_{\vec{s}\, \vec{s}^{\, \prime}} &= \chi_{\mathcal{W}}(m_1) \chi_{\mathcal{W}}(m_2) \left( \mathcal{D} A_{m_1, m_1^\prime} \delta_{m_1^\prime, h_1} \delta_{m_2, m_2^\prime} + \mathcal{D} A_{m_2, m_2^\prime} \delta_{m_2^\prime, h_2} \delta_{m_1, m_1^\prime} \right) \nonumber \\
	&+ \chi_{\mathcal{W}}(m_1) [1 - \chi_{\mathcal{W}}(m_2)] \left( \mathcal{D} A_{m_1, m_1^\prime} \delta_{m_1^\prime, h_1} \delta_{m_2, m_2^\prime} + \frac{\mathcal{D}}{k_{m_2}} A_{m_2, m_2^\prime} \delta_{m_1, m_1^\prime} \right) \nonumber \\
	&+ \chi_{\mathcal{W}}(m_2) [1 - \chi_{\mathcal{W}}(m_1)] \left( \frac{\mathcal{D}}{k_{m_1}} A_{m_1, m_1^\prime} \delta_{m_2, m_2^\prime} + \mathcal{D} A_{m_2, m_2^\prime} \delta_{m_2^\prime, h_2} \delta_{m_1, m_1^\prime} \right) \nonumber \\
	&+  [1 - \chi_{\mathcal{W}}(m_1)] [1 - \chi_{\mathcal{W}}(m_2)] \left( \frac{\mathcal{D}}{k_{m_1}} A_{m_1, m_1^\prime} \delta_{m_2, m_2^\prime} + \frac{\mathcal{D}}{k_{m_2}} A_{m_2, m_2^\prime} \delta_{m_1, m_1^\prime} \right),
\end{align}
\end{widetext}
where $\chi_{\mathcal{W}}$ is the indicator function defined by
\begin{eqnarray}
\chi_{\mathcal{W}}(m) = 
\begin{cases}
  1 & \text{if } m \in \mathcal{W}\,, \\
  0 & \text{if } m \notin \mathcal{W}\,.
\end{cases}
\end{eqnarray}

In an absorbing state, the two walkers are copresent in one of the shared leaf subpopulations.
Therefore, there are $N_s$ absorbing states.
Each element of the $N_T \times N_w$ matrix of transition rates from a transient state to an absorbing state, $R_{\vec{s}\,\vec{s}^{\,\prime}}$, where $\vec{s} = (m_1, m_2) \in \mathcal{B}$ and $\vec{s}^{\,\prime} = (m_1^\prime, m_2^\prime) \in \mathcal{A}$, is given by
\begin{align} \label{eq:R_diff_home}
R_{\vec{s}\, \vec{s}^{\, \prime}} =& \chi_{\mathcal{W}}(m_1) [1 - \chi_{\mathcal{W}}(m_2)] \frac{\mathcal{D}}{k_{m_2}} A_{m_2, m_1^\prime} \delta_{m_1, m_1^\prime} \nonumber \\
      &+ \chi_{\mathcal{W}}(m_2) [1 - \chi_{\mathcal{W}}(m_1)] \frac{\mathcal{D}}{k_{m_1}} A_{m_1, m_2^\prime} \delta_{m_2, m_2^\prime}\,.
\end{align}

The distribution of the initial state of the Markov process, is given by
\begin{align}\label{eq:p0_diff_home}
p_{i, (m_1, m_2)}(0) =& \frac{1}{2} (1 - \delta_{m_1, m_2}) \left( A_{i, m_1} \delta_{m_1, h_1} \delta_{m_2, i} \right. \nonumber \\
     & \left. + A_{i, m_2} \delta_{m_2, h_2} \delta_{m_1, i} \right).
\end{align}
Then, with Eqs.~\eqref{eq:Q_diff_home}, \eqref{eq:R_diff_home}, and \eqref{eq:p0_diff_home} as inputs to our theory, the calculation steps to derive the distribution and CV of IET remain the same.

The CV values obtained from the simulation and the theory for the case in which the two walkers share the home and the case in which they have different homes are shown in Figs.~\ref{fig:work-home_model}(c) and \ref{fig:work-home_model}(d), respectively.
In both cases, the results are similar to those shown in Fig.~\ref{fig:CV_curves}, i.e., the CV is large when $\mathcal{D}/\lambda$ is small, and the theory agrees with the simulation.
The exponential ansatz solution yields slightly smaller values of CV than the numerical simulations, which we also observe in Fig.~\ref{fig:CV_curves}.

\section{\label{sec:discussion}Discussion} 

We have shown that IET distributions are mixtures of exponential distributions for mobile individuals in metapopulation networks.
The results hold true under mild conditions, i.e., for various structures of the metapopulation network, a work-home mobility rule, and higher-order random walks.
Although a mixture of exponential distributions is technically not heavy-tailed, it often approximates heavy-tailed distributions reasonably well over scales \cite{Feldmann1998PerfEval, Okada2020RSOS, Masuda2013bookchap, Papadopoulos2019PRE}.
Therefore, the present results provide a compelling explanation of heavy-tailed IET distributions in human and animal contact data.
Additional mechanisms such as circadian or weekly rhythms \cite{Malmgren2008PNAS} and dynamics of individuals' internal states (e.g., high-activity versus low-activity states) \cite{Reis2020PRE} on top of mobility and metapopulation networks may make IET distributions more smooth and more power-law-like. 

We assumed that two walkers in the same subpopulation meet each other according to a Poisson process.
In other words, the waiting time until they have the next event obeys an exponential distribution if neither walker leaves the current subpopulation.
In fact, this assumption is unnecessary for our results to hold true.
One can assume other types of distributions for the mentioned waiting time to obtain qualitatively the same results.
This is because the result that the IET distribution is a mixture of exponential distributions owes to the fact that the ICT distribution is a mixture of exponential distributions.

The present work bridges statistics of IETs and metapopulation models, which have been extensively but in most cases separately investigated topics in network science and related research fields. 
For example, heavy-tailed IET distributions suppress epidemic spreading in a majority of scenarios \cite{Min2011PRE, Karsai2011PRE, Rocha2011PlosComputBio, Miritello2011PRE, Masuda2013F1000prime, Jo2014PRX, PastorSatorras2015RMF, Masuda2017book}.
The present results allow a new interpretation of these results.
That is, different results of epidemic spreading associated with non-Poissonian IET statistics may owe in large part to whether the structure of the metapopulation network plays a significant role (which would yield heavier-tailed IET distributions) or the population is sufficiently well-mixed (which would yield an exponential IET distribution).
The effects of IET statistics other than their heavy-tailed distributions \cite{Gauvin2018arXiv, Karsai2018book, Masuda2020Book} on epidemic and other dynamical processes may also be due to the underlying metapopulation network.
Reexamining the role of IETs in contagion and other dynamical processes from the viewpoint of metapopulation networks warrants future work.

\begin{acknowledgments}
EFdR acknowledges the support by the Coordena\c{c}\~{a}o de Aperfei\c{c}oamento de Pessoal de N\'{i}vel Superior - Brasil (CAPES) - Finance Code 001.
NM thanks the AFOSR European Office (under Grant no.~FA9550-19-1-7024), the Nakatani Foundation, the Sumitomo Foundation, and Japan Science and Technology Agency (JST) Moonshot R\&D (under grant no.~JPMJMS2021) for the financial support.

\end{acknowledgments}

\appendix

\section{\label{appendix:CV} Coefficient of variation}
The coefficient of variation (CV) is defined as
\begin{eqnarray} \label{eq:CV_def}
\mathrm{CV} = \sqrt{\frac{\langle \tau^2 \rangle}{\langle \tau \rangle^2} - 1}\,,
\end{eqnarray}
where $\langle \cdot \rangle$ is the expectation.
From Eq.~(\ref{eq:general_IET_dist}), we obtain the first and second moments of $\tau$ as
\begin{align} \label{eq:general_first_moment}
\langle \tau \rangle &= -\left. \frac{d \hat{p}(s)}{ds} \right\rvert_{s=0} \nonumber \\
	& = \frac{1}{\lambda}\left( 1 - 2\mathcal{D}\frac{d\hat{\phi}_{\rm c}(0)}{d s}\right)
\end{align}
and
\begin{align} \label{eq:general_second_moment}
\langle \tau^2 \rangle &= \left. \frac{d^2 \hat{p}(s)}{d s^2} \right\rvert_{s=0} \nonumber \\
	& = \frac{2 \mathcal{D}}{\lambda} \frac{d^2 \hat{\phi}_{\rm c}(0)}{ds^2} + \frac{1}{\lambda^2}\left( 1-2\mathcal{D} \frac{d \hat{\phi}_{\rm c}(0)}{ds} \right)^2.
\end{align}

\section{\label{appendix:ansatz_dist}Distribution of interevent times with the exponential ansatz}

In this section, we show that Eq.~(\ref{eq:ansatz_IET_dist}) is a mixture of two exponential distributions.
We first rewrite Eq.~(\ref{eq:ansatz_IET_dist}) as
\begin{align} \label{eq:ansatz_IET_dist_mix_exp}
p(\tau) =& \, C_1 \left(\frac{\lambda+2\mathcal{D}+\alpha-\Delta}{2}\right) e^{-\frac{(\lambda+2\mathcal{D}+\alpha-\Delta)\tau}{2}} \nonumber \\
    & + C_2 \left(\frac{\lambda+2\mathcal{D}+\alpha+\Delta}{2}\right) e^{-\frac{(\lambda+2\mathcal{D}+\alpha+\Delta)\tau}{2}},
\end{align}
where
\begin{eqnarray} \label{eq:C_1}
C_1 = \frac{\lambda(\Delta-\lambda-2\mathcal{D}+\alpha)}{\Delta(\lambda+2\mathcal{D}+\alpha-\Delta)}
\end{eqnarray}
and
\begin{eqnarray} \label{eq:C_2}
C_2 = \frac{\lambda(\Delta+\lambda+2\mathcal{D}-\alpha)}{\Delta(\lambda+2\mathcal{D}+\alpha+\Delta)}\,.
\end{eqnarray}
The two exponents in Eq.~\eqref{eq:ansatz_IET_dist_mix_exp} are positive because 
\begin{eqnarray}
0 < \Delta = \sqrt{(\lambda+2\mathcal{D}+\alpha)^2-4\alpha\lambda} < \lambda+2\mathcal{D}+\alpha\,.\quad
\end{eqnarray}
Because 
\begin{eqnarray}
\Delta - (\lambda+2\mathcal{D}-\alpha)^2 = 8 \lambda \mathcal{D} > 0\,,
\end{eqnarray}
which implies that
\begin{eqnarray}
\Delta > \pm (\lambda + 2 \mathcal{D} - \alpha)\,,
\end{eqnarray}
the weights $C_1$ and $C_2$ are positive.
Equations \eqref{eq:C_1} and \eqref{eq:C_2} also imply that
\begin{eqnarray}
C_1 + C_2 = 1\,.
\end{eqnarray}
Therefore, Eq.~\eqref{eq:ansatz_IET_dist_mix_exp} is a mixture of two exponential distributions.

\section{\label{appendix:transition_rate_matrix}Transition rate matrix}

An element of the transition rate matrix $W$ from a state $\vec{s}=(m,n)$ to a state $\vec{s}^{\,\prime}=(m^\prime,n^\prime)$ is given by
\begin{widetext}
\begin{align} \label{eq:transition_rate_matrix}
W_{\vec{s}\,\vec{s}^{\,\prime}} =& (1-\delta_{mn}) (1-\delta_{m^\prime n^\prime}) \left[ \frac{\mathcal{D}}{k_m} (A_{m m^\prime}\delta_{n n^\prime}+A_{m n^\prime}\delta_{n m^\prime}) + \frac{\mathcal{D}}{k_n} (A_{n n^\prime}\delta_{m m^\prime}+A_{n m^\prime}\delta_{m n^\prime}) \right] \nonumber \\
     & + (1-\delta_{mn}) \delta_{m^\prime n^\prime} \left[ \frac{\mathcal{D}}{k_m} A_{m m^\prime} \delta_{n n^\prime} + \frac{\mathcal{D}}{k_n} A_{n n^\prime} \delta_{m m^\prime} \right].
\end{align}
\end{widetext}
The first term on the right-hand side of Eq.~\eqref{eq:transition_rate_matrix} represents the transitions between transient states.
We define $Q$ as the $N_T \times N_T$ matrix of transition rates between transient states, i.e., $W_{\vec{s}\,\vec{s}^{\,\prime}}=Q_{\vec{s}\,\vec{s}^{\,\prime}}$, where $\vec{s} \in \mathcal{B}$ and $\vec{s}^{\,\prime} \in \mathcal{B}$.
Then,
\begin{align}
Q_{\vec{s}\,\vec{s}^{\,\prime}} =& \frac{\mathcal{D}}{k_m} (A_{m m^\prime}\delta_{n n^\prime}+A_{m n^\prime}\delta_{n m^\prime}) \nonumber \\
      & + \frac{\mathcal{D}}{k_n} (A_{n n^\prime}\delta_{m m^\prime}+A_{n m^\prime}\delta_{m n^\prime})\,.
\end{align}
The second term on the right-hand side of Eq.~\eqref{eq:transition_rate_matrix} represents the transitions from a transient state to an absorbing state.
We define $R$ as the $N_T \times N$ matrix of transition rates from a transient state to an absorbing state, i.e., $W_{\vec{s}\,\vec{s}^{\,\prime}}=R_{\vec{s}\,\vec{s}^{\,\prime}}$, where $\vec{s} \in \mathcal{B}$ and $\vec{s}^{\,\prime} \in \mathcal{A}$.
Then,
\begin{equation}
R_{\vec{s}\,\vec{s}^{\,\prime}} = \frac{\mathcal{D}}{k_m} A_{m m^\prime} \delta_{n m^\prime} + \frac{\mathcal{D}}{k_n} A_{n m^\prime} \delta_{m m^\prime}\,.
\end{equation}
Note that
\begin{eqnarray} \label{eq:sum_W}
\sum_{\vec{s}^{\,\prime}} W_{\vec{s}\,\vec{s}^{\,\prime}} = \sum_{\vec{s}^{\,\prime} \in \mathcal{B}} Q_{\vec{s}\,\vec{s}^{\,\prime}} + \sum_{\vec{s}^{\,\prime} \in \mathcal{A}} R_{\vec{s}\,\vec{s}^{\,\prime}} = 2\mathcal{D}
\end{eqnarray}
for any $\vec{s} \in \mathcal{B}$.
Equation \eqref{eq:sum_W} represents the fact that the system leaves any transient state at rate $2\mathcal{D}$ owing to the movement of each walker, which occurs at rate $\mathcal{D}$.

The master equation for a transient state $\vec{s} = (m,n)$, which is equivalent to Eq.~(\ref{eq:master_pTi}), is given by
\begin{align}\label{eq:master_pTis}
\frac{d p_{i,\vec{s}}^T(t)}{d t} =& \sum_{\vec{s}^{\,\prime} \in \mathcal{B}} \left[ p_{i, \vec{s}^{\,\prime}}^T(t) Q_{\vec{s}^{\,\prime}\,\vec{s}} -  p_{i,\vec{s}}^T(t) Q_{\vec{s}\,\vec{s}^{\,\prime}} \right] \nonumber \\
         & - \sum_{\vec{s}^{\,\prime}\in\mathcal{A}} p_{i,\vec{s}}^T(t) R_{\vec{s}\,\vec{s}^{\,\prime}} \nonumber \\
	=& -2\mathcal{D} p_{i, \vec{s}}(t) + \sum_{\vec{s}^{\,\prime} \in \mathcal{B}} p_{i, \vec{s}^{\,\prime}}(t) Q_{\vec{s}^{\,\prime}\vec{s}}\,.
\end{align}
In Eq.~\eqref{eq:master_pTis}, we used the fact that the sum of transition rates from any transient state to other states is equal to $2\mathcal{D}$, as shown in Eq.~\eqref{eq:sum_W}.

\section{Distribution of inter-copresence times} \label{appendix:dist_ICT}

To derive the ICT distribution, we first note that Eq.~\eqref{eq:sum_W} leads to
\begin{eqnarray} \label{eq:relation_RL}
R \bm 1_N = L \bm 1_{N_T}.
\end{eqnarray}
Then, by combining Eqs.~(\ref{eq:matrix_fpt_abs}), (\ref{eq:stationary_abs}), and \eqref{eq:relation_RL}, we obtain
\begin{align} \label{eq:ICT_dist_derivation}
\phi_{\rm c}(t) &= \bm q^* F^A(t) \bm 1_N \nonumber \\
	&= \bm q^* P^T_0 e^{-Lt} R \bm 1_N \nonumber \\
	&=  e^{-2\mathcal{D}t} \bm q^* P^T_0 (2 \mathcal{D}I-Q) e^{Qt} \bm 1_{N_T}.
\end{align}
By substituting Eq.~(\ref{eq:diagonal_eQt}) into Eq.~\eqref{eq:ICT_dist_derivation}, we find
\begin{align} 
\phi_{\rm c}(t) &= \sum_{j=1}^{N_T} \bm q^* P^T_0 \bm v_j \bm w_j \bm 1_{N_T} (2\mathcal{D}-\gamma_j) e^{-(2\mathcal{D}-\gamma_j)t} \nonumber \\
	&= \sum_{j=1}^{N_T} c_j \alpha_j e^{-\alpha_j t}. \label{eq:ICT_mix_exp_dist_appendix}
\end{align}

In our model, all transient states are reachable from any transient state.
Therefore, $Q$ is irreducible and, by definition, non-negative.
Hence, by the Perron-Frobenius theorem, the spectral radius of $Q$, denoted by $\rho(Q)$, satisfies $\rho(Q) \leq {\displaystyle \max_{\vec{s}}} \sum_{\vec{s}^{\,\prime} \in \mathcal{B}} Q_{\vec{s}\,\vec{s}^{\,\prime}}$.
Using Eq.~\eqref{eq:sum_W}, we obtain $\rho(Q) \leq 2 \mathcal{D}$, which implies that $2\mathcal{D}-\gamma_j \geq 0$, for all $j$.

In the particular case of two subpopulations connected to each other ($N=2$), the number of transient states is $N_T=1$.
Therefore, Eq.~\eqref{eq:ICT_mix_exp_dist_appendix} becomes
\begin{eqnarray}
\phi_c(t) = \alpha_1 e^{-\alpha_1 t},
\end{eqnarray}
which is equivalent to the exponential ansatz.
In this case, there is one transient state and two absorbing states.
Therefore, $Q=0$, $R=\begin{bmatrix} \mathcal{D} & \mathcal{D} \end{bmatrix}$, $L^{-1} = 1/2 \mathcal{D}$, and $\bm 1_{N_T} = 1$.
From Eq.~(\ref{eq:T_matrix}), we obtain
\begin{eqnarray}
T = P_0^T L^{-1} R = \frac{1}{2} \begin{bmatrix} 1 & 1 \\ 1 & 1 \end{bmatrix},
\end{eqnarray}
such that $\bm q^* = \begin{bmatrix} 1/2 & 1/2 \end{bmatrix}$.
Then, we obtain $\alpha_1$ as
\begin{eqnarray}
\alpha_1 = \frac{1}{\bm q^* P_0^T L^{-1} \bm 1_{N_T}} = 2 \mathcal{D}\,.
\end{eqnarray}
Therefore, for the case of two subpopulations, the ICT distribution is given by
\begin{eqnarray}
\phi_c(t) = 2 \mathcal{D} e^{-2\mathcal{D} t}.
\end{eqnarray}

\section{Exact solution for the interevent time distribution} \label{appendix:full_solution}

In this section, we derive the exact solution for the IET distribution.
The probability density with which the two walkers are copresent in subpopulation $j$ after time $t$ given that the last copresence terminated at time $0$ in subpopulation $i$ is given by $f_{ij}^A(t)$, which is given via Eq.~(\ref{eq:matrix_fpt_abs}).
In other words, $f_{ij}^A(t)$ is the probability density of an ICT, $t$, from $(i,i)$ to $(j,j)$.
Note that $f_{ij}^A(t) = \phi_{\rm c}(j,t|i)$ and the normalization is given by $\sum_{j=1}^N \int_{0}^\infty dt \, f^A_{ij}(t) = 1$.
We denote by $p_{ij}(\tau, n)$ the joint distribution of an IET, $\tau$, and the number of copresences, $n$, such that the event has occurred in subpopulation $i$ and the next event occurs in subpopulation $j$.
The normalization is given by $\sum_{j=0}^N \sum_{n=0}^\infty \int_0^\infty d \tau \, p_{ij}(\tau, n) = 1$.
We derive the Laplace transform of $p_{ij}(\tau, n)$ by extending Eq.~(\ref{eq:p(s,n)}) as follows:
\begin{widetext}
\begin{eqnarray} \label{eq:p_{ij}(s, n)}
\hat{p}_{ij}(s, n) = \frac{\lambda (2\mathcal{D})^n}{(s+\lambda+2\mathcal{D})^{n+1}} \sum_{l_1=1}^N \sum_{l_2=1}^N \cdots \sum_{l_{n-1}=1}^N \hat{\phi}_{\rm c}(l_1, s|i) \hat{\phi}_{\rm c}(l_2, s|l_1) \cdots \hat{\phi}_{\rm c}(j, s|l_{n-1})\,.
\end{eqnarray}
\end{widetext}

We define the $N \times N$ matrix $\hat{\Phi}(s)$ by $[\hat{\Phi}(s)]_{ij} = \phi_{\rm c}(j,t|i)$.
Using Eq.~(\ref{eq:matrix_fpt_abs}), we obtain
\begin{eqnarray} \label{eq:Phi(s)}
\hat{\Phi}(s) = P_0^T \left( sI + L \right)^{-1} R\,.
\end{eqnarray}

We also define the $N \times N$ matrix $\hat{\Pi}(s,n)$ by $[\hat{\Pi}(s,n)]_{ij}=\hat{p}_{ij}(s, n)$.
Using Eq.~\eqref{eq:p_{ij}(s, n)}, we obtain
\begin{align} \label{eq:Pi(s)}
\hat{\Pi}(s) & \equiv \sum_{n=0}^\infty \hat{\Pi}(s,n) \nonumber \\
	&= \frac{\lambda}{s+\lambda+2\mathcal{D}} \sum_{n=0}^\infty \left[ \frac{2\mathcal{D}\hat{\Phi}(s)}{s+\lambda+2\mathcal{D}}  \right]^n \nonumber \\
	&= \lambda \left\{ (s+\lambda) I + 2 \mathcal{D} \left[ I - \hat{\Phi}(s) \right] \right\}^{-1}.
\end{align}

We denote by $\hat{g}(s)$ the stationary distribution in the frequency domain of IET weighted by the stationary probability of the initial location of the two copresent walkers.
By combining Eqs.~\eqref{eq:Phi(s)} and \eqref{eq:Pi(s)}, we obtain
\begin{align} \label{eq:g(s)}
\hat{g}(s) &= \bm q^* \hat{\Pi}(s) \bm 1_N \nonumber \\
	&= \lambda \bm q^* \left\{ (s+\lambda)I + 2 \mathcal{D} \left[ I - P_0^T \left( sI + L \right)^{-1} R \right] \right\}^{-1} \bm 1_N,
\end{align}
where $\bm q^*$ is given by Eq.~(\ref{eq:stationary_abs}). 

With Eq.~\eqref{eq:g(s)}, the first and second moments of IET are given by
\begin{eqnarray}\label{eq:<tau>_full}
\langle \tau \rangle = \frac{1}{\lambda} \left( 1 + 2 \mathcal{D} \bm q^* \hat{\Pi}(0) P^T_0 L^{-1} \bm 1_{N_T} \right)
\end{eqnarray}
and
\begin{align}\label{eq:<tau^2>_full}
\langle \tau^2 \rangle =& \frac{4 \mathcal{D}}{\lambda} \bm q^* \hat{\Pi}(0) P^T_0 L^{-2} \bm 1_{N_T} \nonumber \\
        & + \frac{2}{\lambda^2} \bm q^* \left( \hat{\Pi}(0) + 2\mathcal{D} \hat{\Pi}(0) P^T_0 L^{-2} R \right)^2 \bm 1_{N},
\end{align}
respectively, where
\begin{eqnarray}
\hat{\Pi}(0) = \lambda \left[ (\lambda + 2 \mathcal{D}) I - 2\mathcal{D} P_0^T L^{-1} R \right]^{-1}.
\end{eqnarray}
By substituting Eqs.~\eqref{eq:<tau>_full} and \eqref{eq:<tau^2>_full} into Eq.~\eqref{eq:CV_def}, we obtain
\begin{widetext}
\begin{eqnarray}\label{eq:CV_full}
\mathrm{CV} = \sqrt{\frac{4\lambda\mathcal{D} \bm q^* \hat{\Pi}(0) P_0^T L^{-2} \bm 1_{N_T} + 2 \bm q^* \left( \hat{\Pi}(0) + 2 \mathcal{D} \hat{\Pi}(0) P_0^T L^{-2} R \right)^2 \bm 1_N}{\left( 1 + 2 \mathcal{D} \bm q^* \hat{\Pi}(0) P_0^T L^{-1} \bm 1_{N_T} \right)^2} - 1}\,.
\end{eqnarray}
\end{widetext}

\section{\label{appendix:2nd_order_rw}Higher-order random walks}

In this section, we show that our theory holds true when the individuals move according to higher-order random walks.
We focus on second-order random walks and then explain how our theory generalizes to higher orders.

In a second-order random walk, the probability with which a walker visits the next node depends on its current and last visited nodes \cite{Scholtes2014NatComm, Rosvall2014NatComm}.
Therefore, the state of a single walker is defined by a pair of nodes $(m_-, m)$, where $m$ is the currently visited node, and $m_-$ is the node that the walker visited just before arriving in $m$.
It should be noted that $m \neq m_-$ and that we distinguish between $(m_-, m)$ and $ (m, m_-)$.
In other words, the state of each walker is specified by a directed edge.
For example, the sequence of a walker's positions from node $i$ to $j$ and then to $k$ is given by $(i,j) \rightarrow (j,k)$.
Therefore, in a second-order random walk, we can regard the movement of the walkers as a first-order random walk from directed edge $(i,j)$ to directed edge $(j,k)$ instead of between nodes.
We assume in the following text that there are $N$ subpopulations and $M$ undirected edges.
Then, a walker moves among $2M$ directed edges.

The state of the system of two second-order random walkers is defined by a pair of directed edges, one for each walker.
For example, if one walker is currently in subpopulation $m$ and the other walker in subpopulation $n$, we denote the state of the system by $((m_-, m), (n_-, n))$.
Because the walkers are indistinguishable the system has $\overline{N} = 2M(2M+1)/2 = M(2M+1)$ states in total.

We denote the next state of the system by $((m_-^\prime, m^\prime), (n_-^\prime, n^\prime))$ and suppose that the walker at subpopulation $m$ has moved to subpopulation $m_{\rm{new}}$.
Therefore, the new state $((m_-^\prime, m^\prime), (n_-^\prime, n^\prime))$ is equal to $((m, m_{\rm{new}}), (n_-, n))$.
If $m_{\rm{new}} = n$, then the next state (i.e., $((m_-^\prime, m^\prime), (n_-^\prime, n^\prime))$) is an absorbing state.
Otherwise, it is a transient state.

The number of absorbing states with which the two walkers meet at node $i$ is given by $k_i + \binom{k_i}{2} = k_i(k_i+1)/2$, where $k_i$ is the degree of node $i$.
Note that $((j, i), (j, i))$, where $j$ is a neighbor of $i$, is also a valid absorbing state.
Then, the system has $N_A = \sum_{i=1}^N k_i(k_i+1)/2 = M + \frac{1}{2} \sum_{i=1}^N k_i^2$ absorbing states.
The number of transient states is given by $N_T = \overline{N} - N_A = 2 M^2 - \frac{1}{2} \sum_{i=1}^N k_i^2$.

Because we have identified all the transient states, absorbing states, and state transition rules, one is able to define the $N_T \times N_T$ matrix $Q$ of transition rates between transient states and the $N_T \times N_A$ matrix $R$ of transition rates from a transient state to an absorbing state.
Then, the theory that follows is the same as that developed in Section \ref{sec:mixture_solution} and in Appendix \ref{appendix:full_solution}.

It is straightforward to extend the same procedure to the case of directed metapopulation networks and higher-order (i.e., third-order or higher) random walks.
For a third-order random walk, for instance, the state of the system is described by a pair of triples, i.e., $((m_{-2}, m_{-1}, m),(n_{-2}, n_{-1}, n))$, where $m$ is the subpopulation that the first walker currently visits, $m_{-1}$ is the node that the same walker visited just before $m$, and $m_{-2}$ the subpopulation that the walker visited just before $m_{-1}$.
The definitions are analogous for $(n_{-2}, n_{-1}, n)$.

\begin{figure}[ht]
\includegraphics{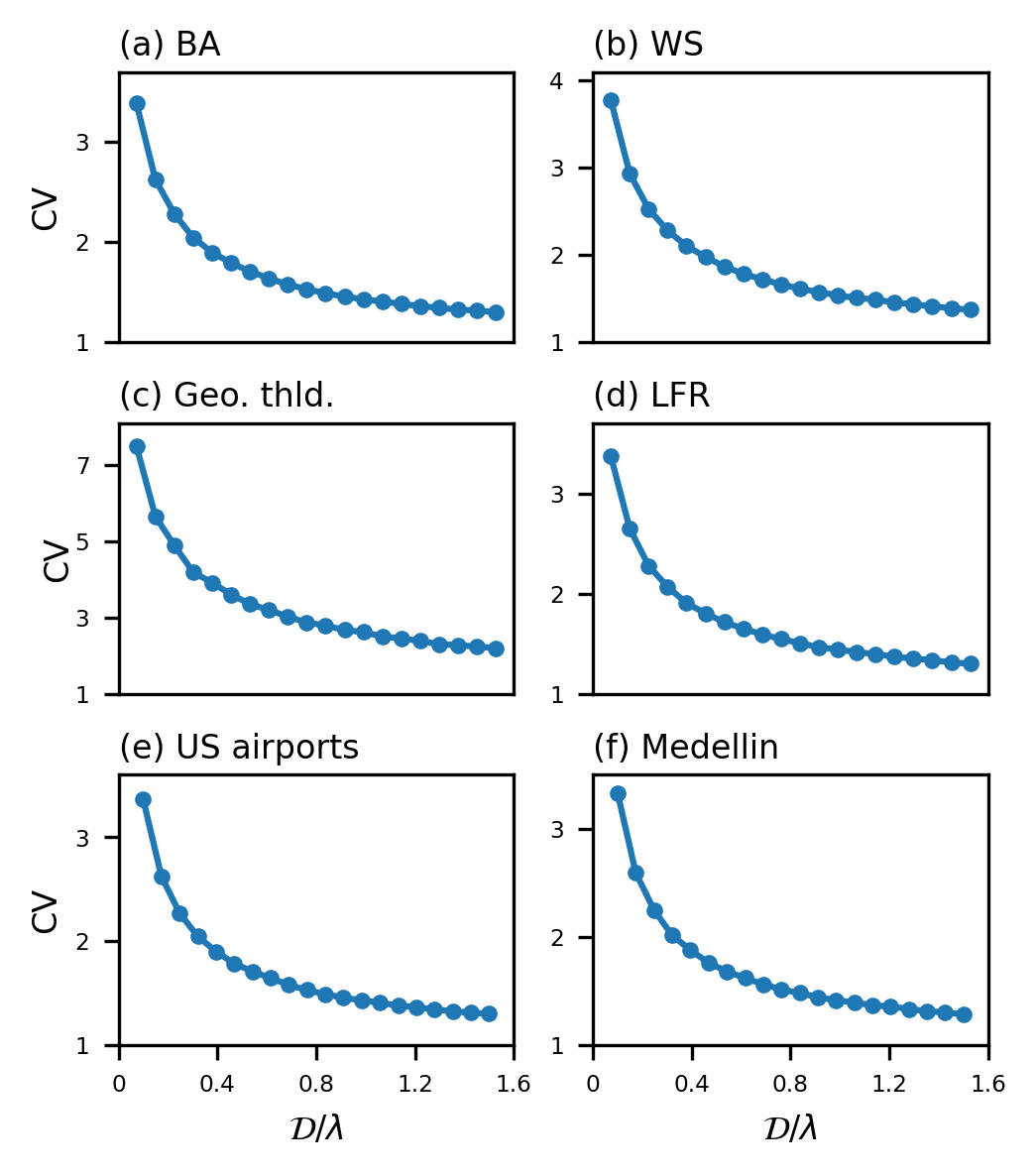}
\caption{\label{fig:CVnbt_curves}
	CV of IET obtained from the simulation of two non-backtracking random walkers for various metapopulation networks.
	(a) Barab{\'a}si-Albert (BA) network with $m=2$.
	(b) Watt-Strogatz (WS) network with $p=0.1$.
	(c) Geographical threshold graph with threshold value $\theta=95$, dimension equal to 2, Euclidean distance metric, and $h(r) = r^{-2}$.
	(d)  Lancichinetti–Fortunato–Radicchi (LFR) model with $\gamma=3$, $\beta=1.5$, $\mu=0.2$, average degree equal to 5, maximum degree equal to 50, and minimum community size equal to 10.
	(e) US airport network.
	(f) Medellin intercity zones.
	In panels (a)--(d), the networks have $N=100$ nodes.
	We set $\lambda=1$.
	}
\end{figure}

As an example of a second-order random walk, we simulated the non-backtracking random walk.
By definition, a non-backtracking random walker on an undirected unweighted network that has moved from $m_-$ to $m$, moves to any of the neighbors except $m_-$ with the equal probability in the next move \cite{Alon2007CommunContempMath,Fitzner2013JStatPhys}.
The CV of IET for two non-backtracking random walkers on various networks is shown in Fig.~\ref{fig:CVnbt_curves}.
The results are similar to those for the simple random walk shown in Fig.~\ref{fig:CV_curves}.
In other words, the CV is substantially larger than 1 and large when $\mathcal{D}/\lambda$ is small.
In general, the CV values for the non-backtracking random walk are somewhat smaller than those for the simple random walk.

%

\end{document}